\documentclass[preprint2]{aastex}
\usepackage{lscape}
\slugcomment{Not to appear in Nonlearned J., 45.}

\shorttitle{Mid\- Infrared in AGN and Starburst}
\shortauthors{Sales;Pastoriza;Riffel}

\begin{document}


\title{Polycyclic Aromatic Hydrocarbon and Emission Line Ratios in Active Galactic Nuclei and Starburst Galaxies.}


\author{Dinalva A. Sales\altaffilmark{1};M. G. Pastoriza\altaffilmark{1,2} and R. Riffel\altaffilmark{1}}
\altaffiltext{1}{Departamento de Astronomia, Universidade Federal do Rio Grande do Sul. Av. Bento Gon\c calves 9500, Porto Alegre, RS, Brazil}
\altaffiltext{2}{Conselho Nacional de Desenvolvimento Cient\' ifico e Tecnol\' ogico, Brazil}
\email{dinalva.aires@ufrgs.br;pastoriza@ufrgs.br;riffel@ufrgs.br}



\begin{abstract}

We study the polycyclic aromatic hydrocarbons (PAH) bands, ionic
emission lines, and mid-infrared continuum properties, in a sample of 171 emission line galaxies 
taken from the literature plus 15 new active galactic nucleus
(AGN) \textit{Spitzer} spectra.  We normalize the spectra at $\lambda$\,=\,23$\mu$m and grouped them 
according to the type of nuclear activity. The continuum 
shape steeply rises for longer wavelengths and can be fitted with a warm blackbody distribution 
of \textit{T} $\sim$ 150\,--\,300K. The brightest PAH spectral bands (6.2, 7.7, 8.6, 11.3, and 12.7\,$\mu$m)
and the forbidden emission lines of [Si{\sc\,ii]} 34.8\,$\mu$m, [Ar{\sc\,ii]} 6.9\,$\mu$m, 
[S{\sc\,iii]} 18.7 and 33.4\,$\mu$m were detected in all the starbursts
 and in $\sim$80\% of the Seyfert~2. Taking under consideration only 
the PAH bands at 7.7$\mu$m, 11.3$\mu$m, and 12.7$\mu$m we find that they are present in $\sim$80\% of the Seyfert~1, 
while only half of this type of activity show the 6.2$\mu$m  and 8.6 $\mu$m PAH bands.
The observed intensity ratios for neutral and ionized PAHs (6.2$\mu$m/7.7$\mu$m $\times$ 
11.3$\mu$m/7.7$\mu$m) were compared to theoretical intensity ratios, showing that AGNs
have higher ionization fraction and larger PAH molecules ($\geq$ 180 carbon atoms) 
than SB galaxies.  The ratio between the ionized (7.7$\mu$m) and the neutral PAH bands 
(8.6$\mu$m and 11.3$\mu$m) are distributed over different ranges for AGNs and SB galaxies, suggesting that these 
ratios could depend on the ionization fraction, as well as on the hardness of the radiation field.
The ratio between the 7.7$\mu$m and 11.3$\mu$m bands is nearly constant with the 
increase of [Ne{\sc\,iii]}15.5$\mu$m/[Ne{\sc\,ii]} 12.8$\mu$m, indicating that the fraction of ionized to 
neutral PAH bands does not depend on the hardness of the 
radiation field. The equivalent width of both PAH features show the same dependence (strongly decreasing) 
with [Ne{\sc\,iii]}/[Ne{\sc\,ii]}, suggesting that the PAH molecules, emitting either ionized 
(7.7$\mu$m) or neutral (11.3$\mu$m) bands, may be destroyed with the increase of the hardness 
of the radiation field.

\end{abstract}





\section{Introduction}

The Mid-infrared (Mid-IR) spectra of galaxies either with active nucleus (AGN) and/or Starburst (SB) show emission features attributed to Polycyclic Aromatic Hydrocarbons (PAH), which can be considered to be originated in very small amorphous carbon dust grains or very large carbon-rich ring molecules \citep[e.g.][]{pugetleger,drainelee84,drainelee01}. The most prominent, well-known PAH emissions, are the 6.2$\mu$m, 7.7$\mu$m, 8.6$\mu$m, 11.2$\mu$m and 12.7$\mu$m bands \citep[e.g.][]{roche91,genzel98,weedman05,buchanan06}. Besides, the Mid-IR spectral region also presents prominent forbidden ionic emission lines, such as [Ne{\sc\,ii]\,}12.8$\mu$m, [Ne{\sc\,v]\,}14.3$\mu$m, [Ne{\sc\,iii]\,}15.5$\mu$m, [S{\sc\,iii]\,}18.7$\mu$m and 33.48$\mu$m, [O{\sc\,iv]\,}25.89$\mu$m and [Si{\sc\,ii]\,}34.8$\mu$m \citep[e.g.][]{sturm02,bernardsalas09}. 

The nature of the ionizing source can be assessed by computing the intensity ratios between forbidden lines such as [Ne{\sc\,ii]\,}12.8$\mu$m, [Ne{\sc\,iii]\,}15.5$\mu$m and [Ne{\sc\,v]\,}14.3$\mu$m \citep[e.g.][]{avoit92}. In addition, the detection of [Ne{\sc\,v]\,} at 14.3$\mu$m and 24.2$\mu$m or [Ne{\sc\,vi]\,} at 7.6$\mu$m, due to ions of high ionization potential, is an evidence of a hard radiation field associated with an AGN \citep{genzel98,sturm00}.

Several diagnostic diagrams based on the line intensity ratios between the brightest  ionic  lines and the PAH features have been proposed to classify the Mid-IR spectra of galaxies according to the degree of activity (AGN, low-ionization nuclear emission-line region - LINER, and SB galaxies). For instance, \citet{genzel98}, using data from the Short Wavelength Spectrometer (SWS) on board of the Infrared Space Observatory (ISO), propose the diagnostic diagram, [O{\sc\,iv]\,}25.9$\mu$m/[Ne{\sc\,ii]\,}12.8$\mu$m versus the strength of the 7.7$\mu$m PAH band to separate  star-forming galaxies from AGNs.

 From the study of the Mid-IR spectra of quiescent spirals, Galactic H{\sc ii} regions, planetary nebulae, and photodissociation regions (PDRs), \citet{galliano06}   found that the ratios between the 6.2$\mu$m/8.6$\mu$m and 7.7$\mu$m/8.6$\mu$m PAH bands do not vary significantly, while the ratios 6.2$\mu$m/11.3$\mu$m and 7.7$\mu$m/11.3$\mu$m
varies over one order of magnitude. They concluded that the properties of the PAHs for different types of galaxies are very similar and that variations of the PAH band ratios are due to the fraction of ionized to neutral PAH molecules. Thus, this ratio could be used to analyze the physical conditions of the molecular gas in the regions where the emission is originated. In addition,
\citet{gordon08} found that the equivalent width (EW) of PAH features, observed in H{\sc ii} regions of the M101 galaxy are correlated with the 
ionization index, and does not depend on the H{\sc ii} regions metallicity.

According to \citet{smith07} the ionized to neutral fraction of PAHs can be estimated using PAH 7.7$\mu$m/11.3$\mu$m ratio, which is sensitive to
the type of nuclear activity, decreasing from SB to Seyfert.  These authors, suggest that this behavior is due to the 
selective destruction of the 7.7$\mu$m PAH molecule with the increasing hardness of the radiation field 
\citep[similar results are found by][]{odowd09,hunt10,kaneda08}.  Such results are supported by 
laboratory experiments, which show that certain PAHs are effectively destroyed by individual UV and X-ray 
photons, and cannot  survive within a few kiloparsecs from an active nucleus, unless they are shielded from 
the AGN's X-ray emission by  absorbing material \citep{bvoit92}.

It is worth mentioning that the above results, concerning the ionization fraction of PAH bands in AGNs are carried out with small samples of this type of object.
Therefore, we use the largest sample of AGNs and SB galaxies available up to date in Spitzer public archive, to study in detail the effect of the hardness of the radiation field on the PAH bands emission, in these kind of galaxies. 
This paper is structured as follows: in Sect. \ref{sample} we describe the sample selection, data reduction process and emission line measurements. In Sect. \ref{results} we present results and discussions on the shape of the continuum spectra, emission line frequency and  diagnostic diagrams of PAH band ratios and emission lines ratios. Sect. \ref{correlacaopahne} we discuss the behavior of the ionization fraction and EW of PAH bands with the hardness of radiation field. Summary and conclucion are presented in Sect. \ref{conclusions}.

\section{The Sample}\label{sample}

We have analyzed a well defined sample of 186 galaxies which have Spitzer spectra available from public archive. Being 83 AGNs taken from \citet{gallimore10}, 22 SB from \citep{brandl06}, 59 H{\sc ii} and LINER from \citep{smith07} and 7 H{\sc ii} regions of M101 galaxy from \citep{gordon08}, plus 15 AGNs Spitzer spectra published here for the first time. Details on the reduction process of the 15 new objects are given below.

\subsection{Data Reduction}\label{data}

\begin{figure*}
\centering
\includegraphics[scale=0.45,angle=0]{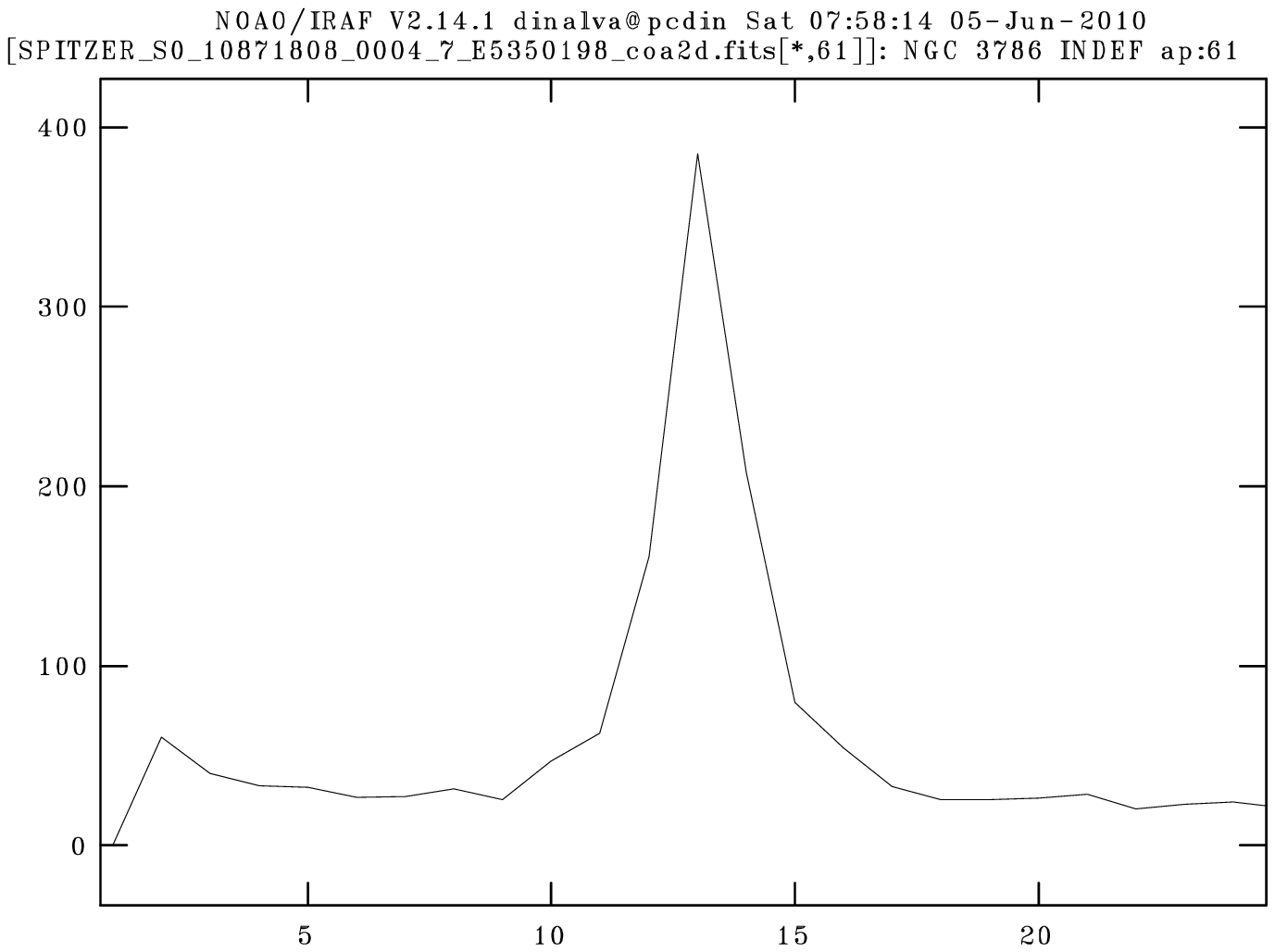}
\includegraphics[scale=0.45,angle=0]{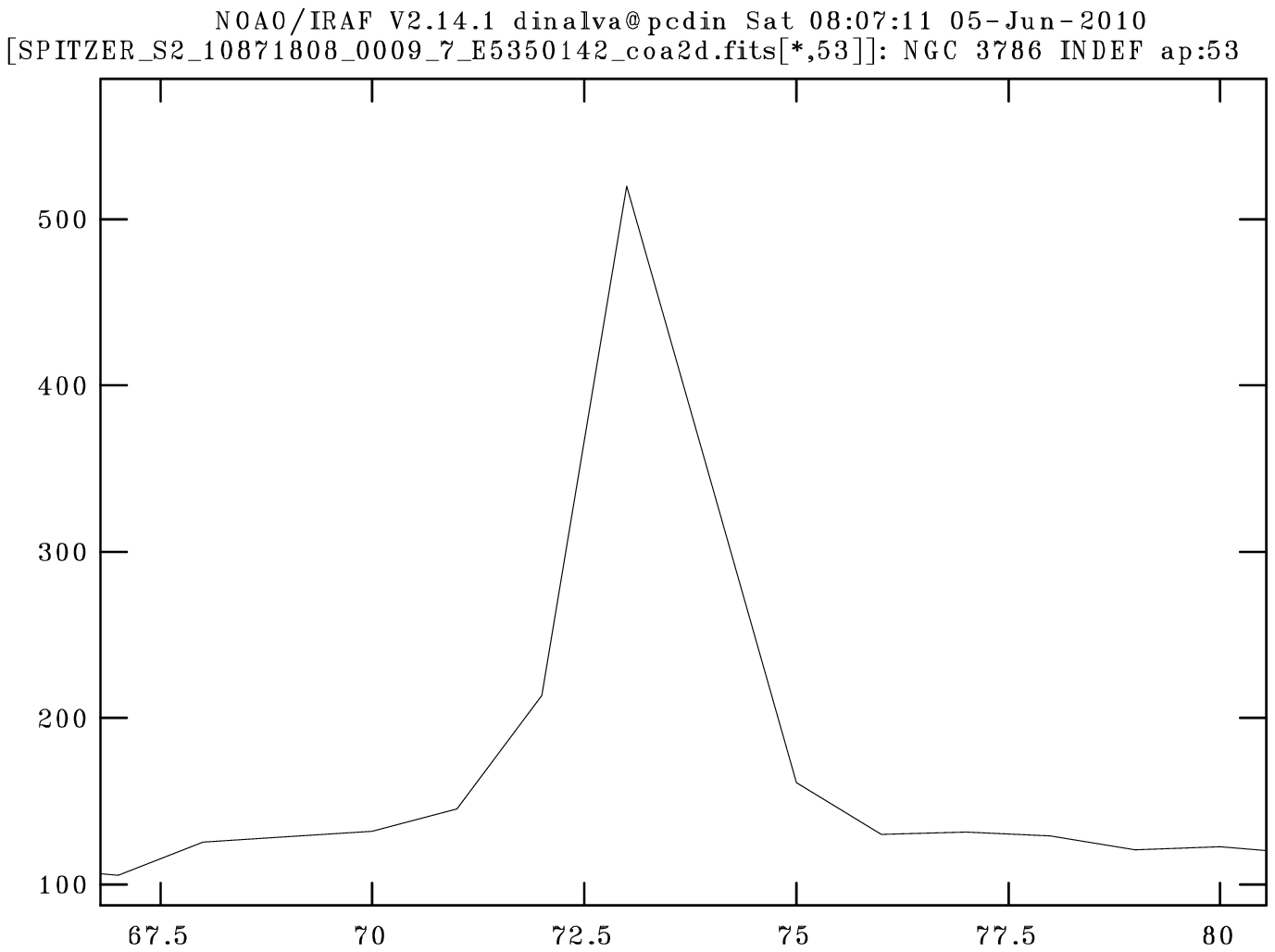}
\caption{Spatial emission profile of NGC3786 for the SL (left) and LL (right) modules.}
\label{profile}
\end{figure*}

The 15 new spectra of AGN galaxies (4 Sy 1 and 11 Sy 2)  were obtained from the Spitzer public archive. Observations were performed  with  the Infrared Spectrograph \citep[IRS;][]{houck04} using the Short-Low (SL) and Long-Low (LL) modules, covering the interval between 5$\mu$m and 36$\mu$m with a resolving power of 64 - 128. The raw data were processed using  the Spitzer {\sc pipeline} version 17.2\footnote{Available at: http://ssc.spitzer.caltech.edu/spitzerdataarchives/}. The nuclear spectra were extracted using the Spitzer IRS Custom Extractor - SPICE\footnote{Available at: http://ssc.spitzer.caltech.edu/postbcd/spice.html.}. Default point-source extractions were selected for all objects. Only a few  sources show a noticeable step in flux between the overlapping LL and LS module spectra, which results from the extended circumnuclear emission  contribution  to the size  SL module slit of 3$^{''}$.7 x 57$^{''}$ and  LL module slit of  10$^{''}$.7 x 168$^{''}$. Although the  observed  regions of almost all of the galaxies are large  in both modules the AGN light dominates the spectra. Fig. \ref{profile} shows the emission  spatial profile of NGC3786. Note that the luminosity falls off to almost zero for $\approx$ 3 pixels from the center of the nucleus. For this galaxy,  3 pixels  correspond to $\approx$ 1 kpc, adopting a radial velocity of 2678 km sec$^{-1}$ and a Hubble constant of 74 km sec$^{-1}$Mpc$^{-1}$.

The general parameters for the objects listed in Table \ref{tabgeral} were taken from NASA/IPAC Extragalactic Database (NED). All the spectra were 
corrected for redshift. The rest-frame spectra were grouped according to their activity, and are shown in Fig. \ref{allnorm}.

\subsection{Emission Line Measurements}\label{decomposition}

For the 15 new AGNs listed in the Table \ref{tabgeral} and SB taken from \citet{brandl06}, low resolution IRS spectra have been decomposed using the  {\sc pahfit}\footnote{Source code and documentation for {\sc pahfit} are available in http://tir.astro.utoledo.edu/jdsmith/research/pahfit.php} code, described in detail by \citet{smith07}. In short, they assume that the IRS  spectra are composed by dust continuum, starlight, prominent emission lines, individual and blended PAH emission bands, and that the light is attenuated by extinction due to silicate grains.

The default input parameters of {\sc pahfit} are: (i) the infrared emission from a blackbody with T= 5000\,K; (ii) different weights for thermal dust continuum components represented by the temperatures 35, 40, 50, 90, 135, 200, 300K; (iii) spectral line features, including the pure rotational lines of the molecular hydrogen; (iv) dust features are represented by individual and blended Drude profiles, which are theoretical frequency profiles for a classical damped harmonic oscillator and are the best choice to model PAH emission. For more details see \citet{smith07}.

With the above constraints we were able to isolate the PAH features and measure their fluxes, as well as  those of  the ionized gas emission lines. The fluxes and the EW are listed in Tables \ref{fluxpah} to \ref{fluxionic}. We checked the consistency of the results with the {\sc liner} code \citep{poggeowen93} that integrates the flux of a Gaussian function fitted to the line, obtaining 
similar results in both procedures. The observed IRS spectra of Mrk334, a Sy 1 galaxy, is shown in Fig. \ref{pahfit}, where we also plot the detailed components of the  {\sc pahfit} spectral decomposition  model. Clearly,  the model reproduces well the observed spectrum.

\section{Results and Discussion}\label{results}

\subsection{New spectra}

\begin{figure*}[ht]
%\begin{minipage}[b]{0.5\linewidth}
\centering
\includegraphics[scale=0.5,angle=-90]{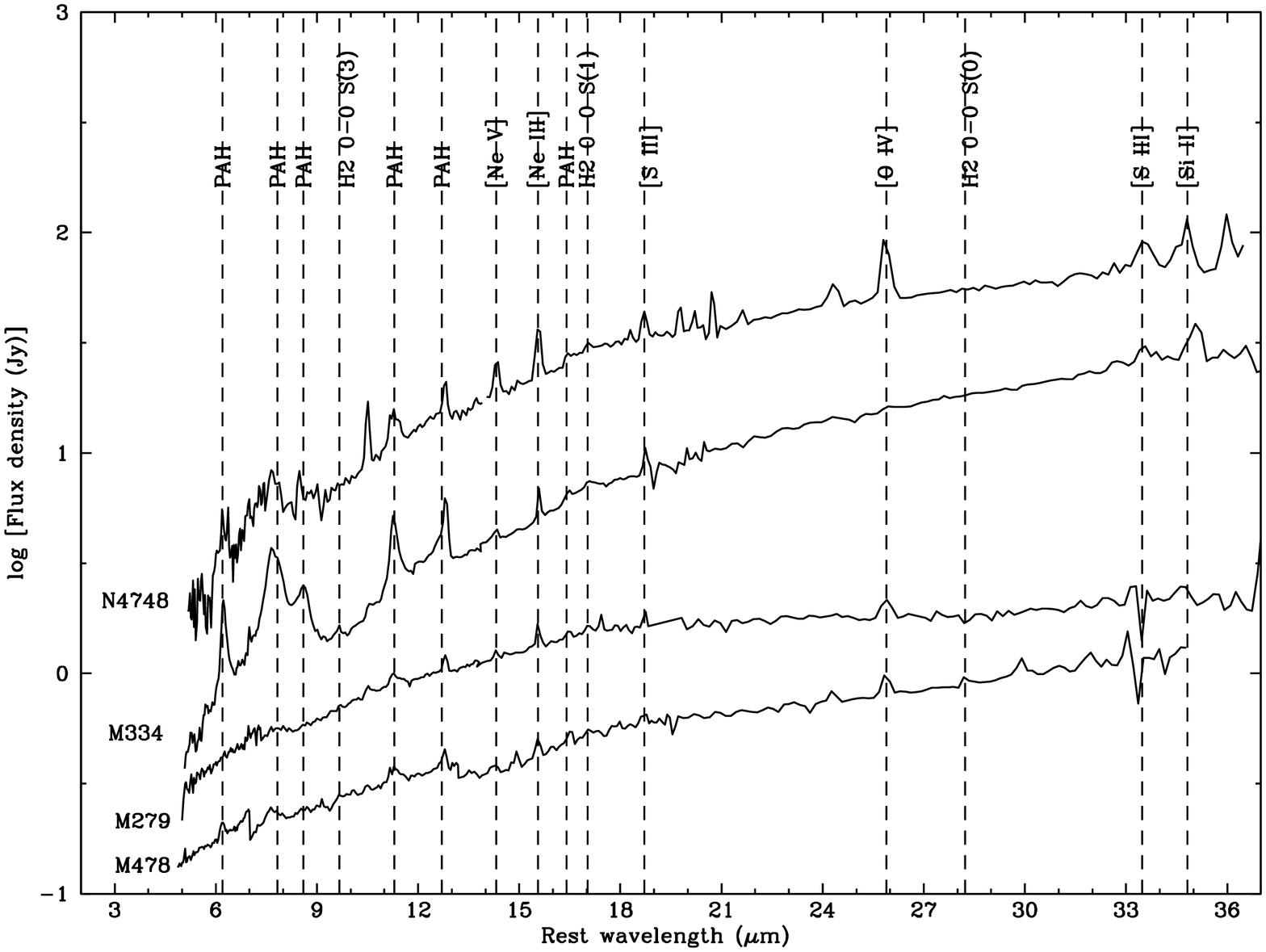}
\includegraphics[scale=0.5,angle=-90]{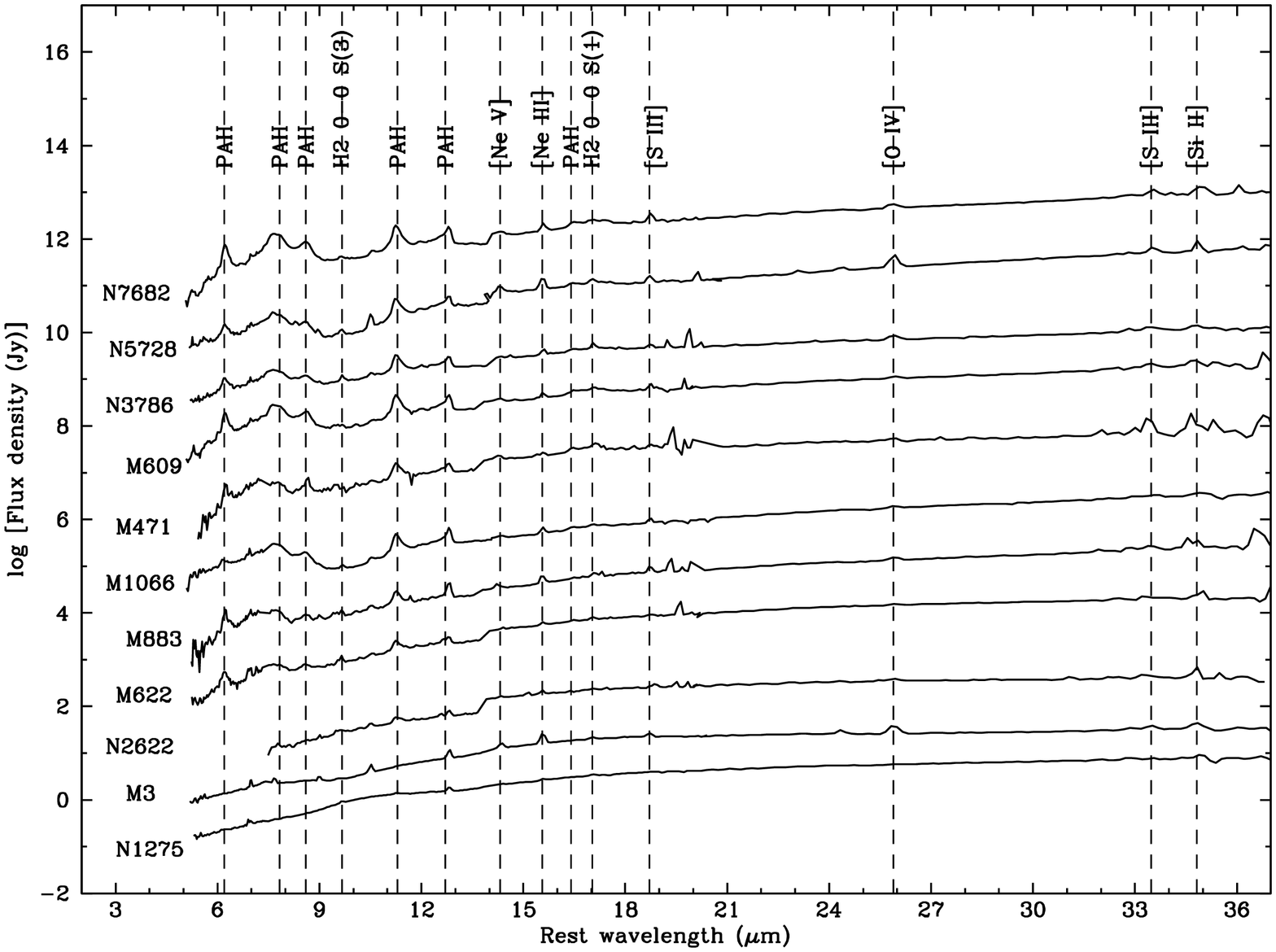}
\caption{Calibrated spectra of the Sy 1 (top) and Sy 2 (bottom) galaxies. All spectra were ordered according to their shapes from a steeper spectrum (top) to a flatter one (bottom). Some emission lines are identified. All spectra were normalized at 23$\mu$m.}
\label{allnorm}
%\end{minipage}
\end{figure*}

The main goal of this section is to characterize the continuum shapes and spectral features observed in the 15 new objects listed in Table \ref{tabgeral}, with respect to  the type of nuclear activity (Sy 1 and Sy 2). Therefore, we normalize to unity the continuum of all spectra at $\lambda$ = 23 $\mu$m and grouped them according to nuclear activity. For each AGN class,  data were sorted according to spectral shape, from bluest to reddest, top to bottom of Fig. \ref{allnorm}, respectively.  One can clearly see that the majority of spectra have a continuum shape increasing for longer wavelengths, in agreement with results found by \citet{weedman05,buchanan06,deo09,burtscher09,wu09,baum}. This continuum range can be fitted with a warm blackbody distribution 
of T $\sim$ 150 --- 300\,K. Interestingly, a similar trend in the continuum of AGNs is also observed in the near-infrared region \citep{riffel06,riffel09}.

We compare the spectroscopic properties of the 15 new objects with those of the \citet{buchanan06} groups. We found that the spectra of the Sy 2 galaxies: NGC3786, NGC5728, NGC7682, Mrk471, Mrk609, Mrk622, Mrk883, Mrk1066, Mrk883 and the Sy 1 Mrk334 and NGC4748, have the same characteristic as their group 1 (strong PAH emission and red continuum, see Fig \ref{allnorm}). On the other hand the Sy 2, NGC 1275, NGC2622, Mrk3 and Sy 1  Mrk 478 can be classified as group 3 (power-law continuum and weak PAH features), while only the Sy 1 Mrk279 galaxy show a broken continuum and silicate emission at 9.8$\mu$m (group 2). In summary, for our and \citet{buchanan06} samples the Mid-IR spectra for most of Sy 2 galaxies have same spectroscopic characteristics as those of group 1 ($>$ 50$\%$).

\subsection{PAH and Emission Line Frequency}\label{freqEmLin}

\begin{figure*}
\centering
\includegraphics[scale=0.5,angle=-90]{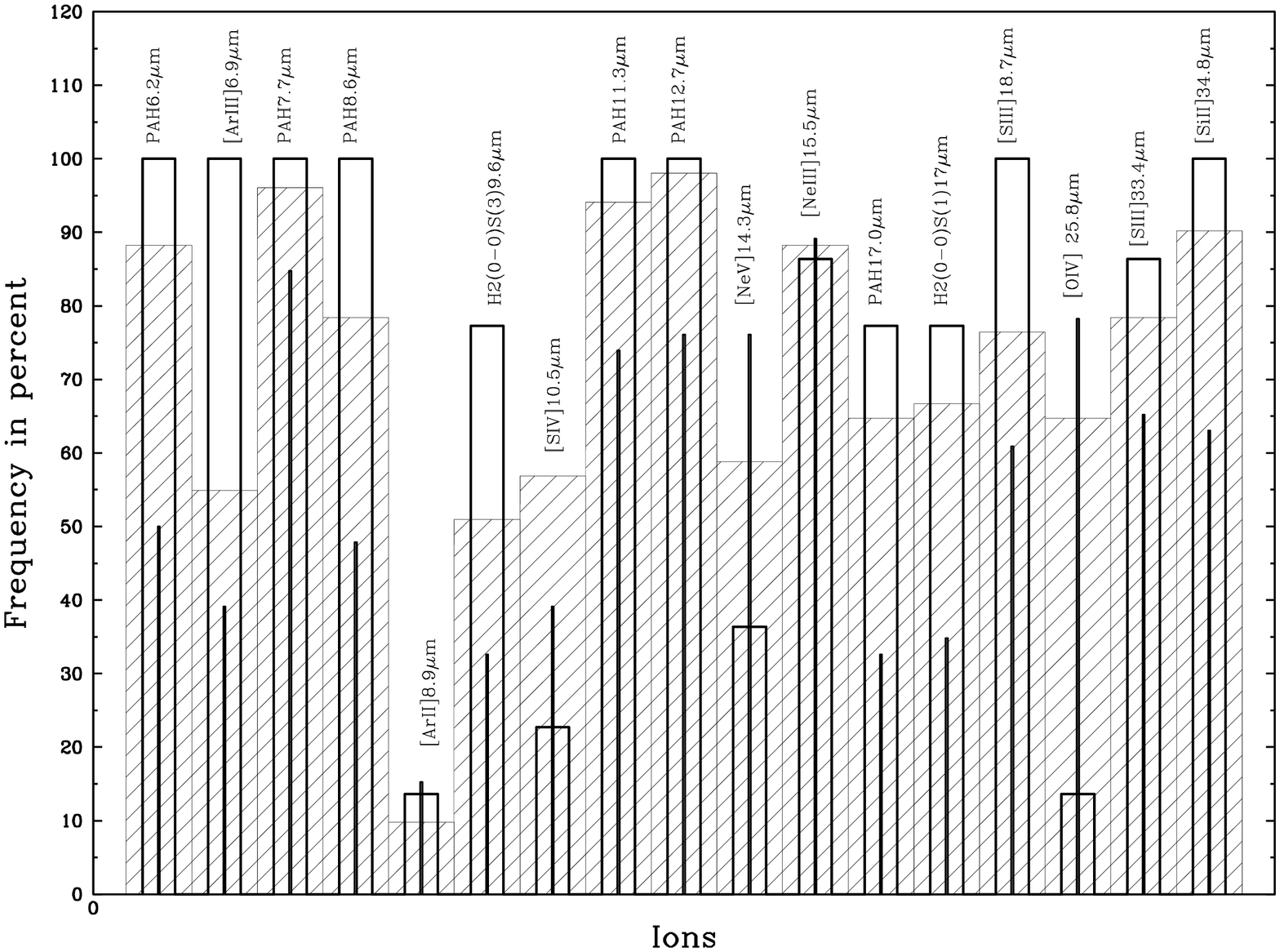}
\caption{Frequency histogram of most common Mid-IR emission lines (in percent). SB, Sy~2 and Sy~1 are represented by empty and shaded bars, and solid lines, respectively.}
\label{hist}
\end{figure*}

\begin{figure*}[ht]
\centering
%\begin{tabular}{ll}
\includegraphics[scale=0.5,angle=0]{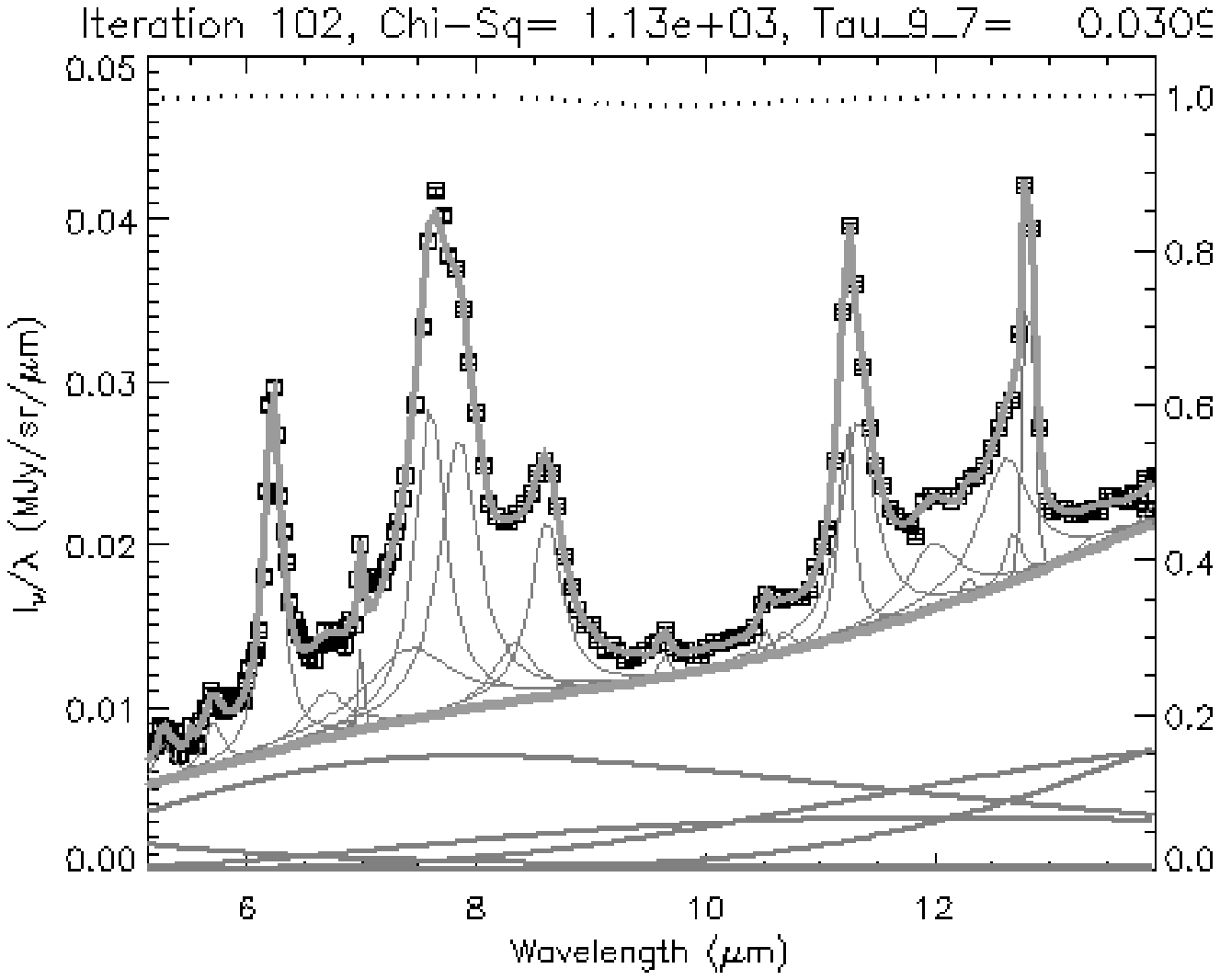}%\tabularnewline
\includegraphics[scale=0.5,angle=0]{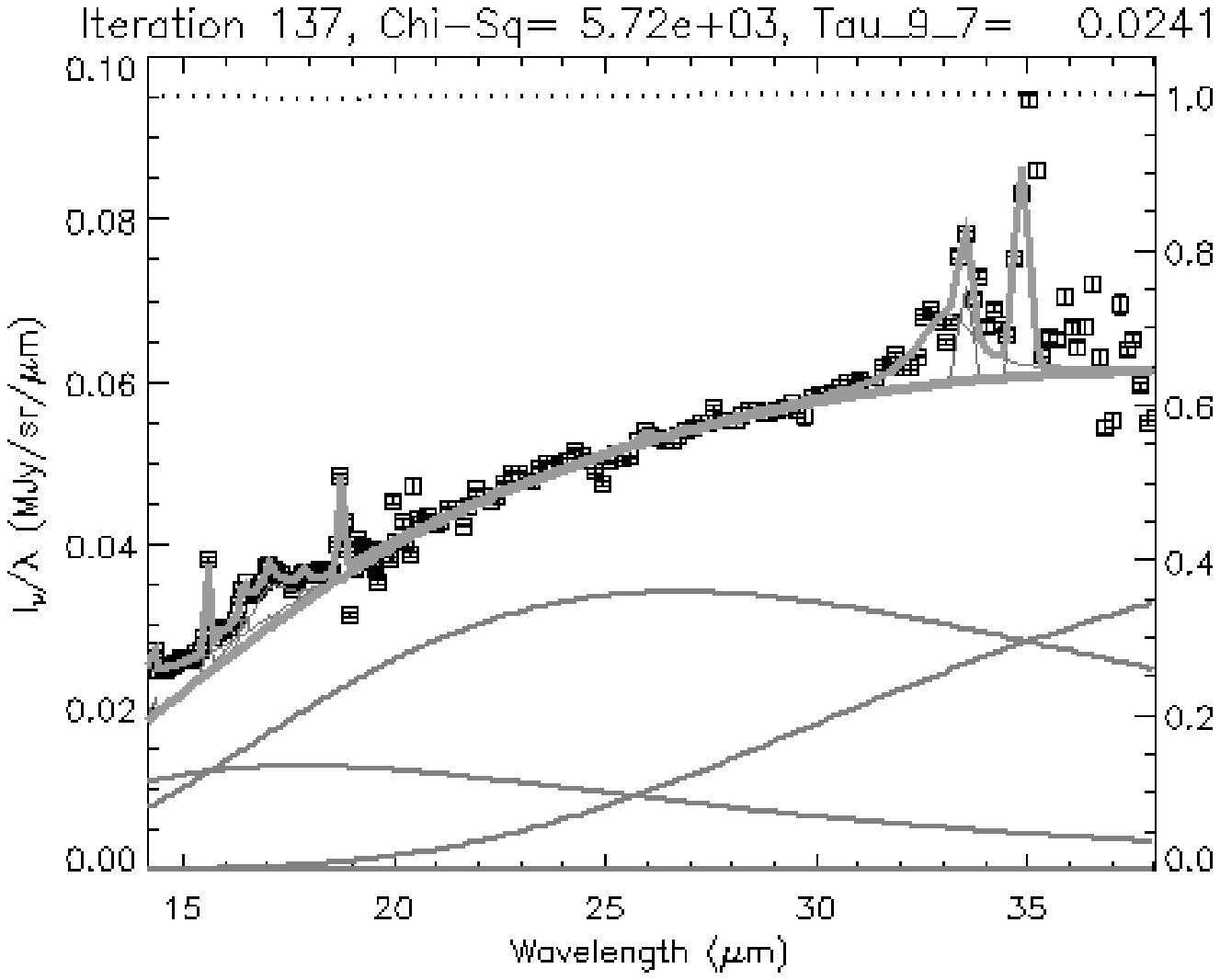}
\caption{Detailed decomposition of Mrk334 from 5 to 37$\mu$m using {\sc pahfit} code. Solid black lines represent thermal dust continuum components. Measurements are represented by squares, uncertainties are plotted as vertical error-bars, which are smaller than the symbol size.  Dotted black line indicates mixed extinction components.  Continuous lines (gray) represent the best fit model with individual PAH features and emission lines, respectively. }
\label{pahfit}
\end{figure*}

Fig.~\ref{hist} shows a histogram of the commonest emission features present in each type of activity (note that we have used the 186 galaxies spectra).
It is clear from this figure that the  PAH features are present in almost all the Sy 1, Sy 2 and SB galaxies, as well as the H$_{2}$ molecules at 9.6$\mu$m and 17$\mu$m. In addition, the most frequent ionic lines are: [Si{\sc\,ii]} 34.8$\mu$m, [Ar{\sc\,ii]} 6.9$\mu$m, [S{\sc\,iii]} 18.7 and 33.4$\mu$m, [S{\sc\,iv]} 10.5$\mu$m, [Ne{\sc\,iii]} 15.53$\mu$m and [Ne{\sc\,v]} 14.3$\mu$m.

All studied SB galaxies show  PAH bands and the following forbidden emission lines: [Si{\sc\,ii]\,}34.8$\mu$m, [Ar{\sc\,ii]\,}6.9$\mu$m, [S{\sc\,iii]\,}18.7 and 33.4$\mu$m. In addition, nearly 80$\%$ show molecular H$_{2}$ lines. Similarly, $\geq$80$\%$ of the Sy 2 show PAH features, and the same ionic lines detected in SB galaxies.  However, we have found that PAH lines in Sy 1 galaxies behave differently from the other types of nuclear activity. The PAH bands at 7.7$\mu$m, 11.3$\mu$m and 12.7$\mu$m are present in 80\% of the galaxies, but 
only 50\% of them show the PAH lines at 6.2$\mu$m and 8.6$\mu$m.  Nevertheless,  the high ionization lines, such as [Ne{\sc\,v]\,}14.3$\mu$m and [O{\sc\,iv]\,}25.8$\mu$m are more common in Sy 1.

\subsection{Diagnostic Diagram of PAH bands: grain sizes and ionization fraction}\label{pahbands}

Numerical studies of the vibrational energy distribution of PAH molecules show that the relative strengths 
of the 3.3$\mu$m, 6.2$\mu$m, 7.7$\mu$m, 8.6$\mu$m and 11.3$\mu$m PAH features depend on grain size and on the charging conditions \citep{drainelee01}. 
Small PAHs radiate strongly at 6.2$\mu$m and 7.7$\mu$m, while large PAHs emit mostly at longer wavelengths \citep{schutte93,draineli07,tielens08}.  Another important conclusion  of \citet{drainelee01}   is   that neutral PAHs have higher values  of 11.3$\mu$m/7.7$\mu$m ratio, while ionized PAHs have lower ratios. On the other hand, the 6.2$\mu$m/7.7$\mu$m ratio  decreases with the increasing number of carbon atoms of the PAH molecules,  Fig. 16 from \citet{drainelee01} shows how the relative strengths of these emission vary depending on the size and charge state of PAH molecules.
\citet{odowd09} used this diagram to analyze these PAH ratios for a sample formed by galaxies dominated by star formation, galaxies with significant stellar and AGN component and  AGN-dominated galaxies. They found a weak trend in the direction of constant ionization fraction with changing grain size. In addition, they suggest that the presence of an AGN component is correlated with the reduction in the 7.7$\mu$m/11.3$\mu$m ratio indicating that smaller PAH grains (number of carbon atoms $<$180) are destroyed in AGNs.

\begin{figure*}[ht]
%\begin{minipage}[b]{0.7\linewidth}
\centering
\includegraphics[scale=0.6,angle=-90]{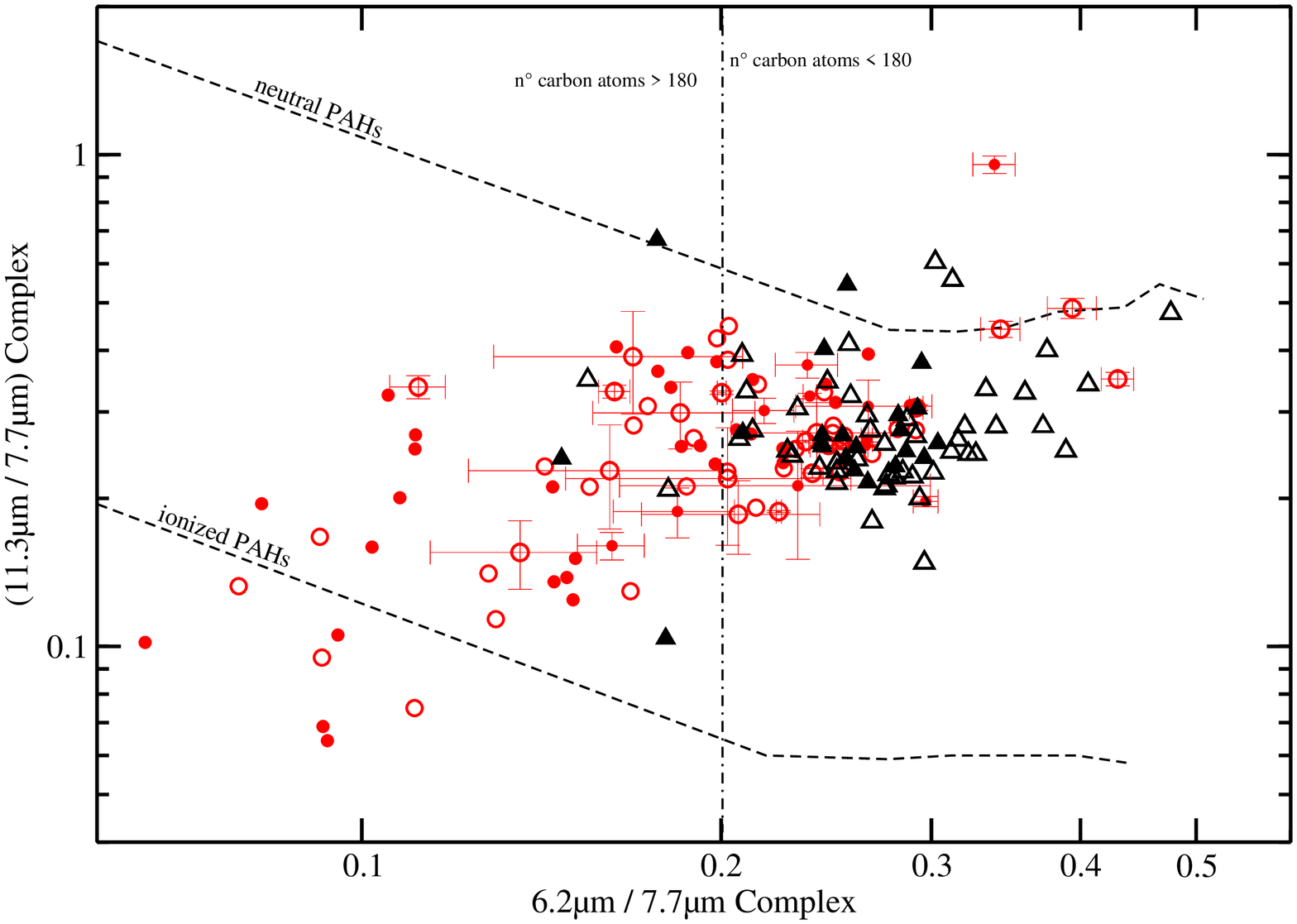}
\centering
\caption{Diagnostic diagram of 6.2$\mu$m/7.7$\mu$m $\times$ 11.3$\mu$m/7.7$\mu$m PAH ratios. The long dash line represents theoretical intensities of neutral to ionized PAHs \citep{drainelee01}. Dashed-dotted line shows the position of molecules formed by 180 carbon atmos. Empty triangles are H{\sc ii} and LINER objects taken from \citet{smith07} and \citet{gordon08}, full triangles are SB taken from \citet{brandl06}, full and empty circles are Sy 1 and Sy 2 of our and \citet{gallimore10} sample. Error bars are shown only for the 15 new AGNs, similar error were found for the remaing objects.}
\label{pahs}
%\end{minipage}
\end{figure*}

We present in Fig. \ref{pahs} the theoretical intensity ratios, for neutral and ionized PAHs (6.2$\mu$m/7.7$\mu$m $\times$ 11.3$\mu$m/7.7$\mu$m), from  \citet{drainelee01}. 
The authors calculated these quantities assuming   grains   illuminated either by the \citet{mathis83} spectrum or a blackbody with T= 3 x 10$^{4}$K varying the number of carbon atoms in each molecule. We overplot in this diagram the observed PAH emission line ratios of our sample. Clearly,  the observed ratios are located  between neutral and ionized PAH theoretical lines, in agreement with  previous studies \citep{drainelee01,odowd09}. 

The above results suggest that the observed PAH bands in both, AGNs and low ionization  objects (SB and LINERS), are formed by an appropriate mixture of PAH molecules with different sizes and adequate neutral to ionized fraction.  High ionization objects (AGNs) present a tendency for molecules $>$ 180  carbon atoms (6.2$\mu$m/7.7$\mu$m $<$ 0.2), on the other hand most of low ionization objects are biased towards molecules with less than 180 carbon atoms.  The latter are typically located in the region 0.2 $<$6.2$\mu$m/7.7$\mu$m $<$ 0.4 (Fig. \ref{pahs}). Furthermore, Seyfert galaxies appear to be located near to region populated by ionized PAH molecules, while the majority of SB, H{\sc ii}, and LINER are near to the line representing the neutral PAHs.  Moreover, these results are consistent with a picture that AGN have higher ionization fraction and larger PAH molecules than SB galaxies.

Another way to assess the ionization fraction of the PAH molecules is using the following diagnostic 
diagram  6.2$\mu$m/11.3$\mu$m $\times$ 7.7$\mu$m/11.3$\mu$m. Such diagram was used by \citet{galliano06}, \citet{galliano08} 
and \citet{odowd09}, they demonstrate that there is a good correlation between both ratios. 
However, we would like to emphasize the fact that both samples were mostly composed by SB galaxies.  In addition, these ratios are 
controlled by the fraction of ionized to neutral PAHs \citep{allamandola99}.

\begin{figure*}[ht]
%\begin{minipage}[b]{0.7\linewidth}
\centering
\includegraphics[scale=0.4,angle=-90]{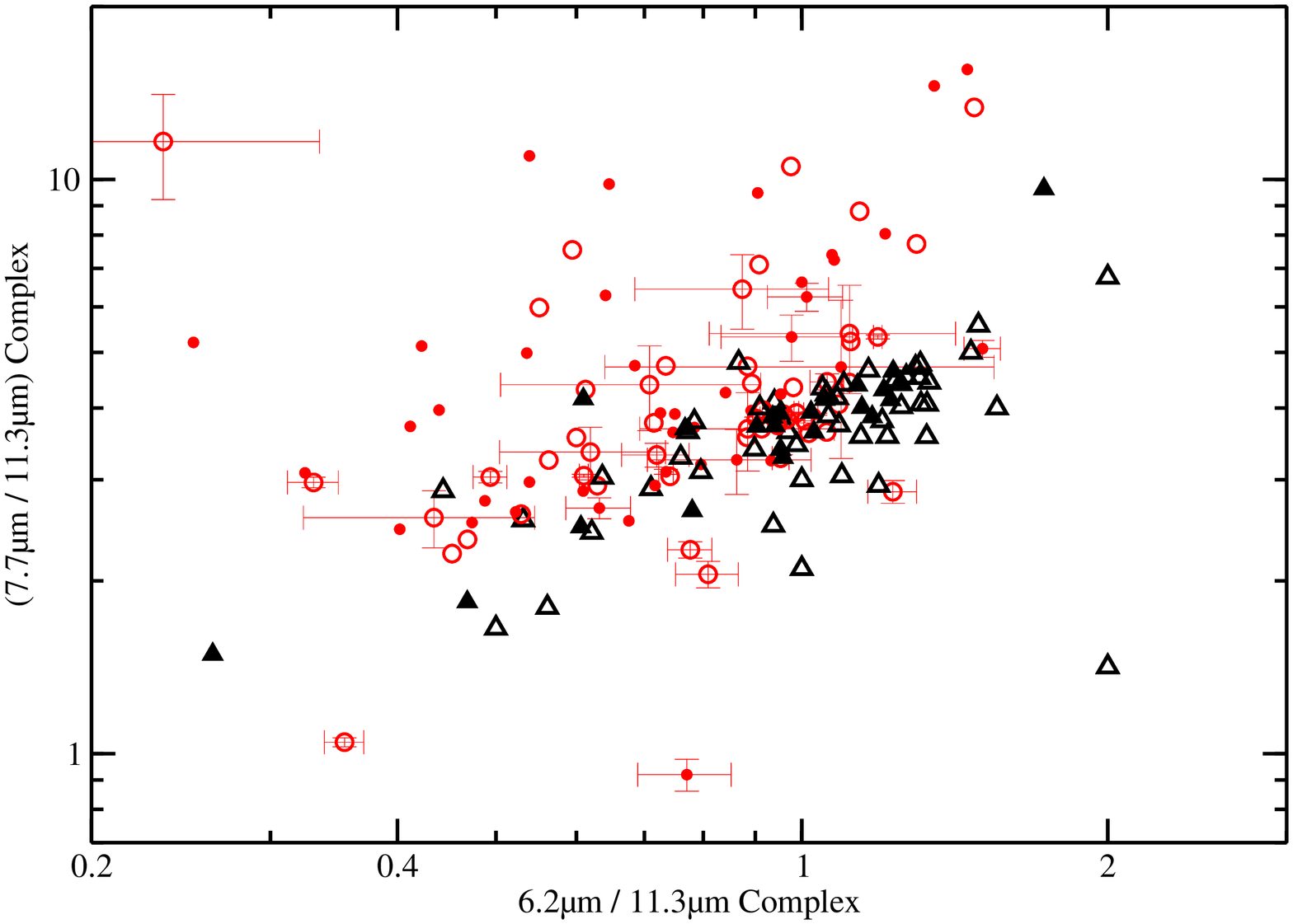}
\includegraphics[scale=0.4,angle=-90]{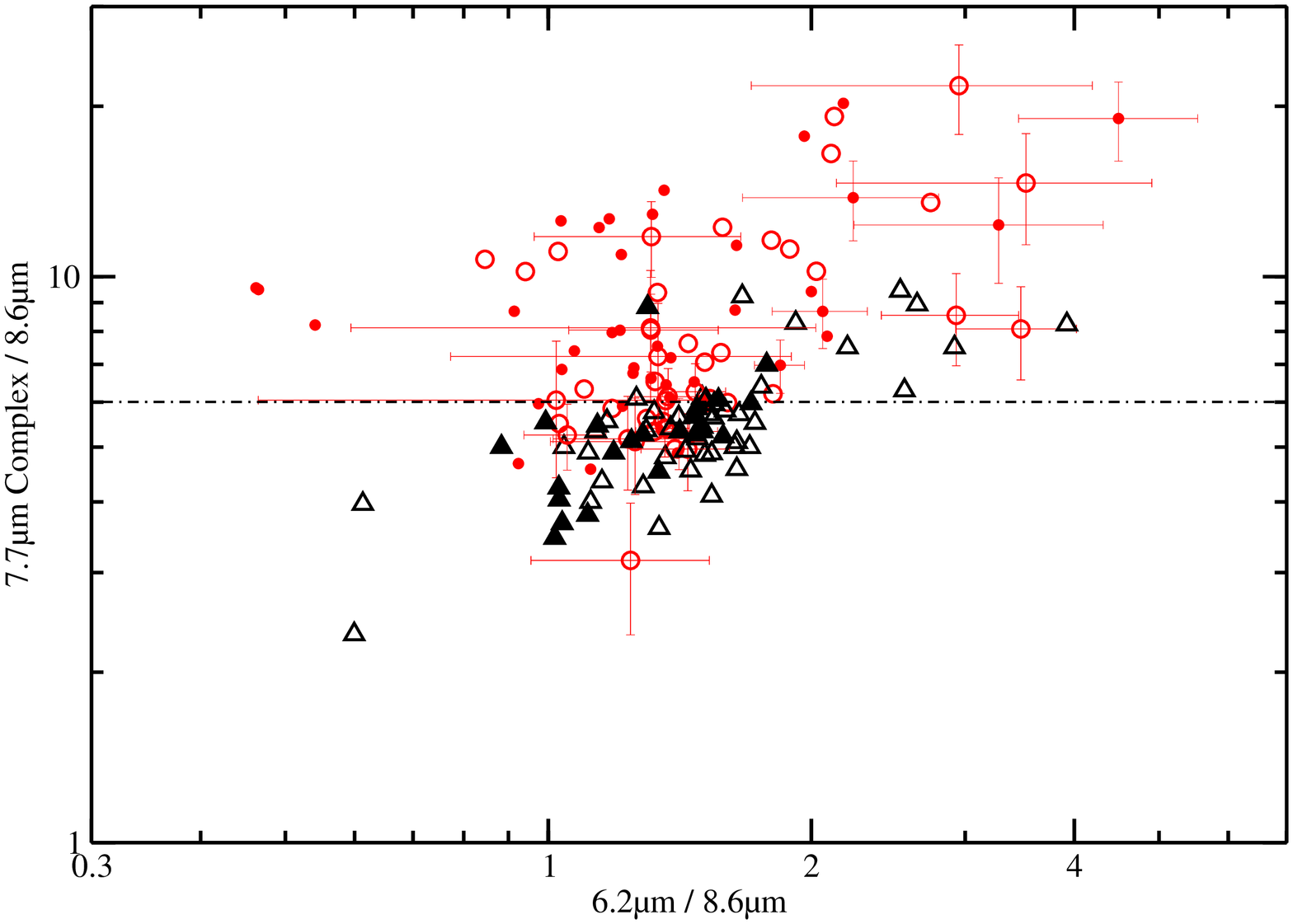}
\centering
\caption{Diagnostic diagram of the relative strengths of the neutral (8.6$\mu$m and 11.3$\mu$m) compared to ionized (6.2$\mu$m and 7.7$\mu$m) PAH bands. In the top, 6.2$\mu$m/11.3$\mu$m $\times$ 7.7$\mu$m/11.3$\mu$m PAHs. Empty triangles are H{\sc ii} and LINER objects taken from \citet{smith07} and \citet{gordon08}, full triangles are SB taken from \citet{brandl06}, full and empty circles are Sy 1 and Sy 2 of our and \citet{gallimore10} sample. Error bars are shown only for the 15 new AGNs, similar error were found for the remaing objects. Dotted line separate high and low ionization objects.}
\label{pahspahs}
%\end{minipage}
\end{figure*}

In Fig \ref{pahspahs} (top) we show these diagram for our sample. Clearly, if we only consider the SB galaxies our results are 
consistent with the correlation described above. By including the AGNs (Sy 1 and Sy 2),  an unexpected large dispersion is observed for both type of activity. Alternatively, we also test
the diagram 6.2$\mu$m/8.6$\mu$m $\times$ 7.7$\mu$m/8.6$\mu$m involving ionized (7.7$\mu$m and 6.2$\mu$m) and neutral (8.6$\mu$m) PAH molecules. 
This diagram is shown in Fig.\ref{pahspahs} (bottom), where a similar dispersion is observed. Interestingly, in both diagnostic diagrams the ratios in the
vertical axes tend to separate the AGNs (Sy 1 and Sy 2) from SB galaxies. This is even clearer in the diagram involving the   7.7$\mu$m/8.6$\mu$m ratio, 
leading us to the conclusion that  7.7$\mu$m/8.6$\mu$m $\geq$ 6 for AGNs (dotted line) and smaller for low ionization objects.

Synthetic spectra in the 6.2$\mu$m -- 9.0$\mu$m region, for large symmetric PAHs cations, anions and
neutral, have been computed by \citet{bauschlicher08}. The trends, in
the band position and intensite, is a function of the molecule size, charge and geometry.
Large PAH cations and anions ($>$ 110 carbon atmos) produce prominente
bands at 7.7$\mu$m and 8.5$\mu$m, the intensite ratio 7.7$\mu$m/8.6$\mu$m increase with the
PAH sizes. Thus, we interpret the separation in activity types (SB, Sy 1 and Sy 2), along the vertical
axis in both diagrams of Fig. \ref{pahspahs}, as due to the fact that in average the emitting
PAH molecules in Seyferts are larger than those of SB. In addition, we suggest
that the high values for the ratios 7.7$\mu$m/8.6$\mu$m and 7.7$\mu$m/11.3$\mu$m is due to the fact
that almost all the molecules are ionized. In summary, we point out that both diagrams
(Figs. \ref{pahs} and \ref{pahspahs}) clearly shows that the ionization fraction and
molecules size increase from SB to AGNs. However, it is worth mentioning that, no separation
between Seyfert type is observed.

Although, the four ratios depends on the PAHs ionization fractions, 
ratios  involving 6.2$\mu$m band have the same range of variation for both types of nuclear activity. 
This can be associated with the fact that the 6.2$\mu$m PAH emission
is due to small molecules \citep{bauschlicher08}, which probably are destroyed by
the hard radiation field of Seyfert galaxies.

\subsection{Diagnostic Diagram for Emission Line Objects}\label{emissionlines}

As shown in Sect.~\ref{freqEmLin} (Fig.~\ref{hist}), forbidden ionic emission lines are observed in SB and active galaxies. In the first case, the gas is ionized by massive hot stars, while in the latter the ionization is due to a non-thermal continuum. Moreover, the emission cloud in AGNs are
located in the narrow line regions (NLR) \citep{osterbrock}.

The Mid-IR emission from dusty NLR of AGNs has been explored by \citet{groves06}.  Their models give a direct connection between the dust emission and the line emission from the photoionized gas. These authors also show  that the emission line ratios [Ne{\sc\,v]}/[Ne{\sc\,ii]} $\times$ [Ne{\sc\,iii]}/[Ne{\sc\,ii]} can be used to separate AGNs from SBs. In addition, \citet{thornley00} using starburst evolutionary models, which incorporate new stellar atmosphere models for massive stars, found that [Ne{\sc\,iii]}/[Ne{\sc\,ii]} is sensitive to hardness of the radiation field.

\begin{figure*}[ht]
\centering
\includegraphics[scale=0.6,angle=-90]{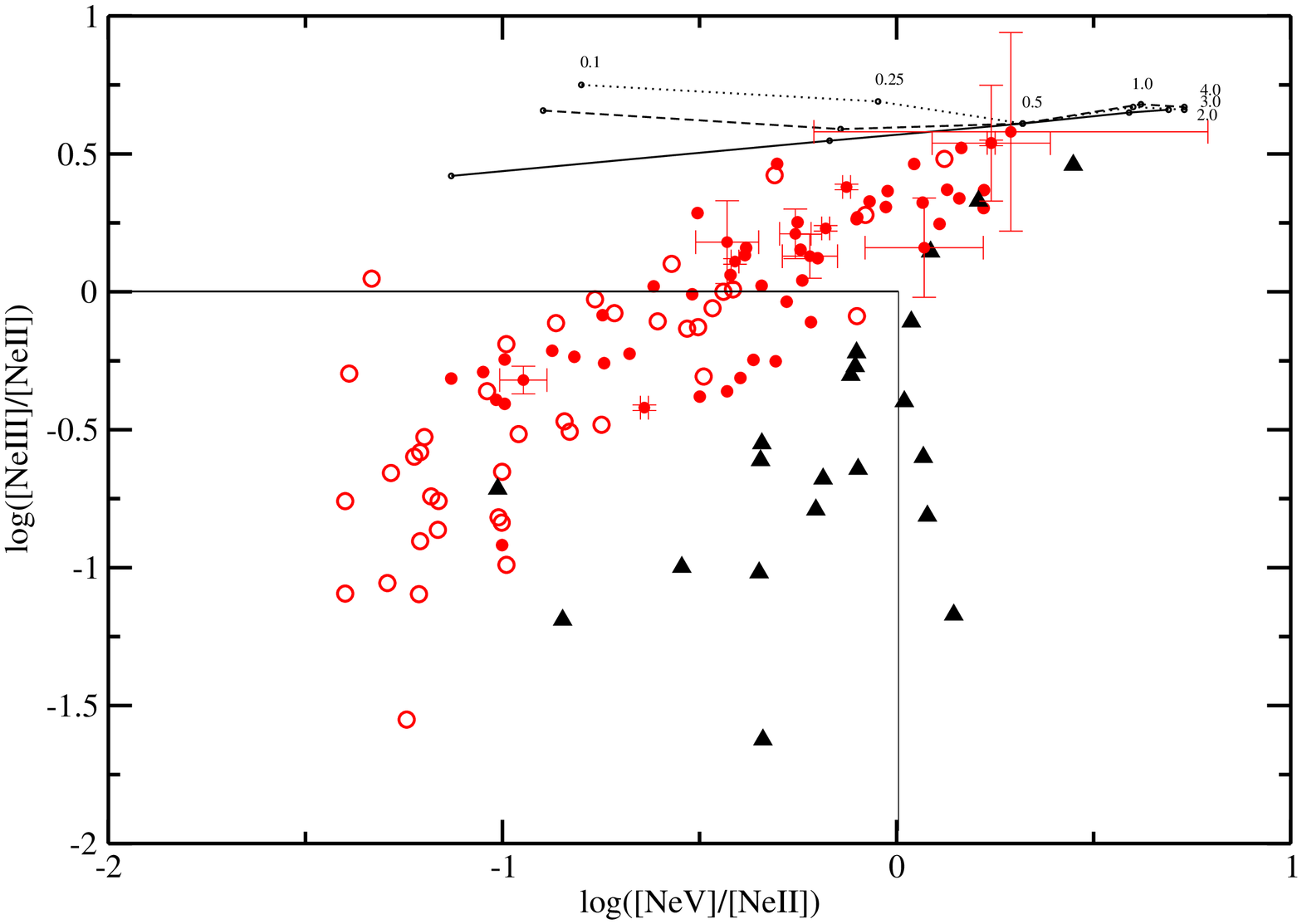}%\tabularnewline
\caption{Diagnostic diagram of [Ne{\sc\,v]\,}14.3$\mu$m/[Ne{\sc\,ii]\,}12.8$\mu$m $\times$ [Ne{\sc\,iii]\,}15.5$\mu$m/[Ne{\sc\,ii]\,}12.8$\mu$m  ratios. The direction of increasing ionization parameter for dusty NLR models with the total pressure of: $\textit{P}_{tot}$/k \,$\simeq 10^6$ (dot curve), $\simeq 10^7$ (dashed curve) and $\simeq 10^8$ (solid curve). Full triangles are SB taken from \citet{brandl06}, full and empty circles are Sy 1 and Sy 2 of our and \citet{gallimore10} sample. Error bars are shown only for the 15 new AGNs, similar error were found for the remaing objects.}
\label{nene}
\end{figure*}

In Fig.~\ref{nene} we show the diagnostic diagram [Ne{\sc\,v]}14.3$\mu$m/[Ne{\sc\,ii]}12.8$\mu$m $\times$ [Ne{\sc\,iii]}15.5$\mu$m/  [Ne{\sc\,ii]}12.8$\mu$m for our sample. We also overplot the NLR models for  different  ionization and pressure parameters taken from \citet{groves06}. Clearly, all the observations have values lower than those predicted by the models. 
In this diagram, AGNs are closely correlated, while the SB galaxies show a large scattering.

The [Ne{\sc\,v]}14.3$\mu$m has a high ionization potential of 97.1 eV and is often detected in spectra of AGNs \citep[e.g.][]{sturm02,weedman05} however is rarely  detected in SB galaxies \citep{brandl06,bernardsalas09}. The lack of correlation for SB galaxies in this diagram, thus, may be due to a strong contamination of [Ne{\sc\,v]} with PAH 14.19$\mu$m emission or to the absence of the former line in the SB galaxies spectra. In addition, we call the attention to the fact that the deblending of these lines is very difficult in the low resolution Spitzer spectra.

It is also clear in this figure that Sy~1 galaxies have [Ne{\sc\,iii]\,}/[Ne{\sc\,ii]\,} $\geq$ 0, while SB and Sy~2 have this ratio $\leq$ 0 (solid line in Fig.~\ref{nene}). This suggests that both line ratios are sensitive to the hardness of the radiation field, confirming the result of \citet{thornley00}. 
 The fact that Sy~2 galaxies are spread over the lower left of the Fig.~\ref{nene} can be associated to \citet{groves06} predictions, which states that  
starburst contribution would move the points in that direction.

It is worth mentioning that there are 3 SB galaxies located in the top-right side of Fig.~\ref{nene} (NGC1097, NGC4676 and NGC520). The former,
was reclassified as Sy 1 by \citet{thaisa97}, while the two latter are strongly interacting sources and may have the these emission lines enhanced \citep{read98}. Thus we suggest that further, more accurate studies (i.e. with higher spectral resolution spectra), are necessary to better explain the  position of the SB galaxies in the diagram.

\section{Behavior of the Ionization Fraction and EW of PAH Molecules with the Hardness of Radiation Field}\label{correlacaopahne}

\citet{smith07} have studied the correlation between  the intensity ratio of the strongest PAH bands  7.7$\mu$m/11.3$\mu$m   with the hardness of the radiation field indicator, [Ne{\sc\,iii]\,}/[Ne{\sc\,ii]\,}. They have found that galaxies with H{\sc ii} regions or starburst optical spectra exhibit a nearly constant 7.7$\mu$m/11.3$\mu$m   across the whole range of radiation hardness, including low-metallicity galaxies, which have [Ne{\sc\,iii]\,}/[Ne{\sc\,ii]\,} $<$ 2. In contrast, the AGNs are located below the H{\sc ii}-type galaxies and their PAH ratio falls rapidly with 
increasing radiation field. This result was confirmed by \citet{odowd09} through the analysis of PAH emission bands of a sample of 92 typical star forming galaxies. They interpreted the decrease in the PAHs band ratio for AGNs as due to a selective destruction of the ionized 7.7$\mu$m PAH molecules.

\begin{figure*}[ht]
\centering
\includegraphics[scale=0.6,angle=-90]{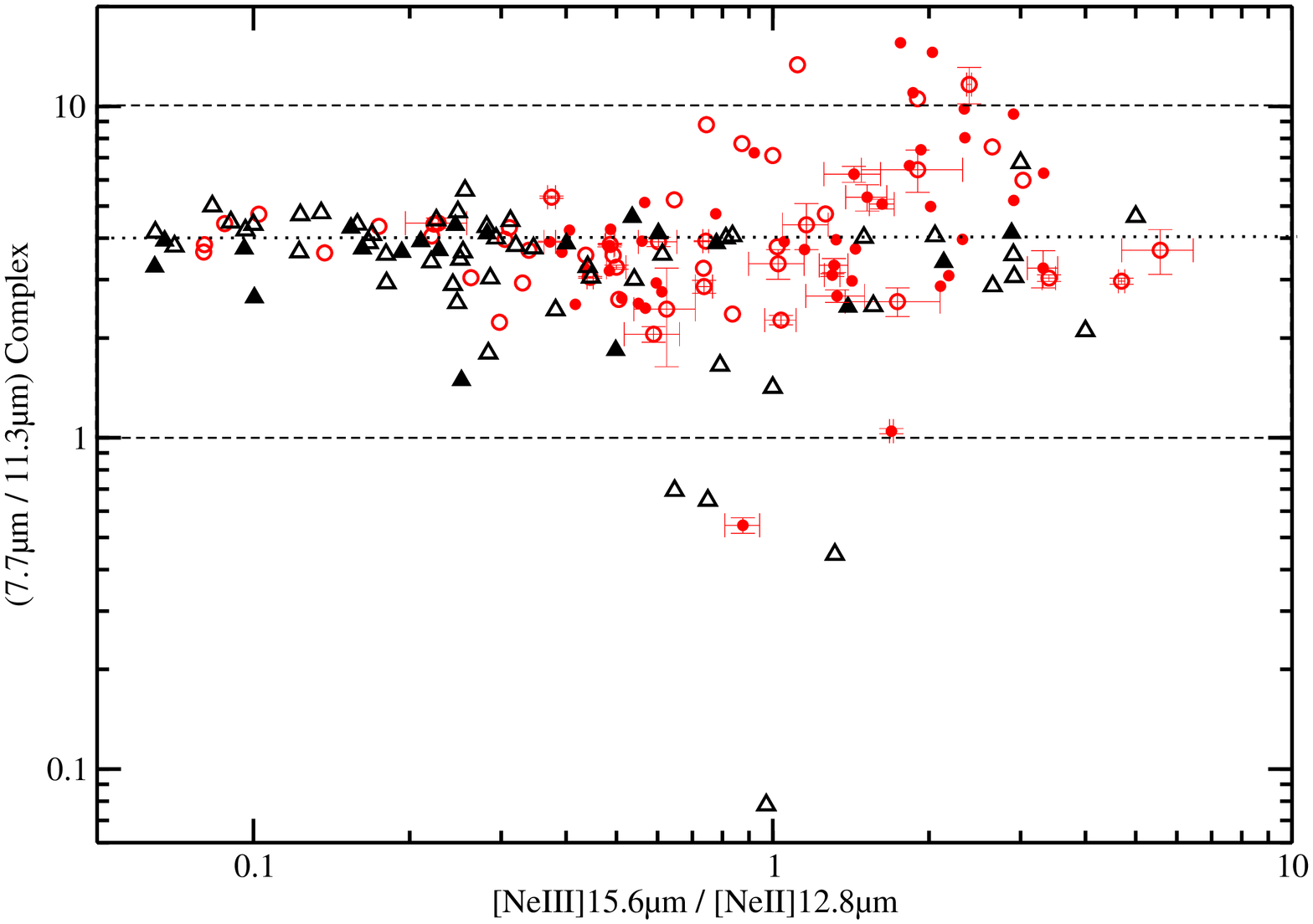}%\tabularnewline
\caption{Diagnostic diagram between an indicator of the hardness of the radiation field, [Ne{\sc\,iii]\,}15.5/[Ne{\sc\,ii]\,}12.8$\mu$m $\times$ 7.7$\mu$m/11.3$\mu$m PAH emission features. Empty triangles are H{\sc ii} and LINER objects taken from \citet{smith07} and \citet{gordon08}, full triangles are SB taken from \citet{brandl06}, full and empty circles are Sy 1 and Sy 2 of our and \citet{gallimore10} sample. Error bars are shown only for the 15 new AGNs, similar error were found for the remaing objects.}
\label{pahne}
\end{figure*}

The variation of 7.7$\mu$m/11.3$\mu$m PAH bands, observed in our sample (186 objects), with respect to the  hardness of the radiation field, is shown in Fig. \ref{pahne}. It is clear from  this figure,  that the fraction of ionized to neutral PAH bands does not depend on the hardness of the radiation field   in contrast with the previous results  of \citet{smith07} and \citet{odowd09}. We interpret this  behavior as due to the fact that the strengths of both PAH bands would have the same dependence on the radiation field. 

Besides, this diagram show a tendency in the sense that SB galaxies have [Ne{\sc\,iii]\,}/[Ne{\sc\,ii]\,} $<$ 0.4, while for Sy 1 and Sy 2 [Ne{\sc\,iii]\,}/[Ne{\sc\,ii]\,} $>$ 0.4. 
In addition, Sy 1 are biased towards location on the upper right corner.
Some of the Sy 2 galaxies are located in the SB region, which suggest that these galaxies may have a circumnuclear starburst.
No clear separation between Sy 1 and Sy 2 is observed for [Ne{\sc\,iii]\,}/[Ne{\sc\,ii]\,} $>$ 0.4.

\begin{figure*}[ht]
\centering
\includegraphics[scale=0.6]{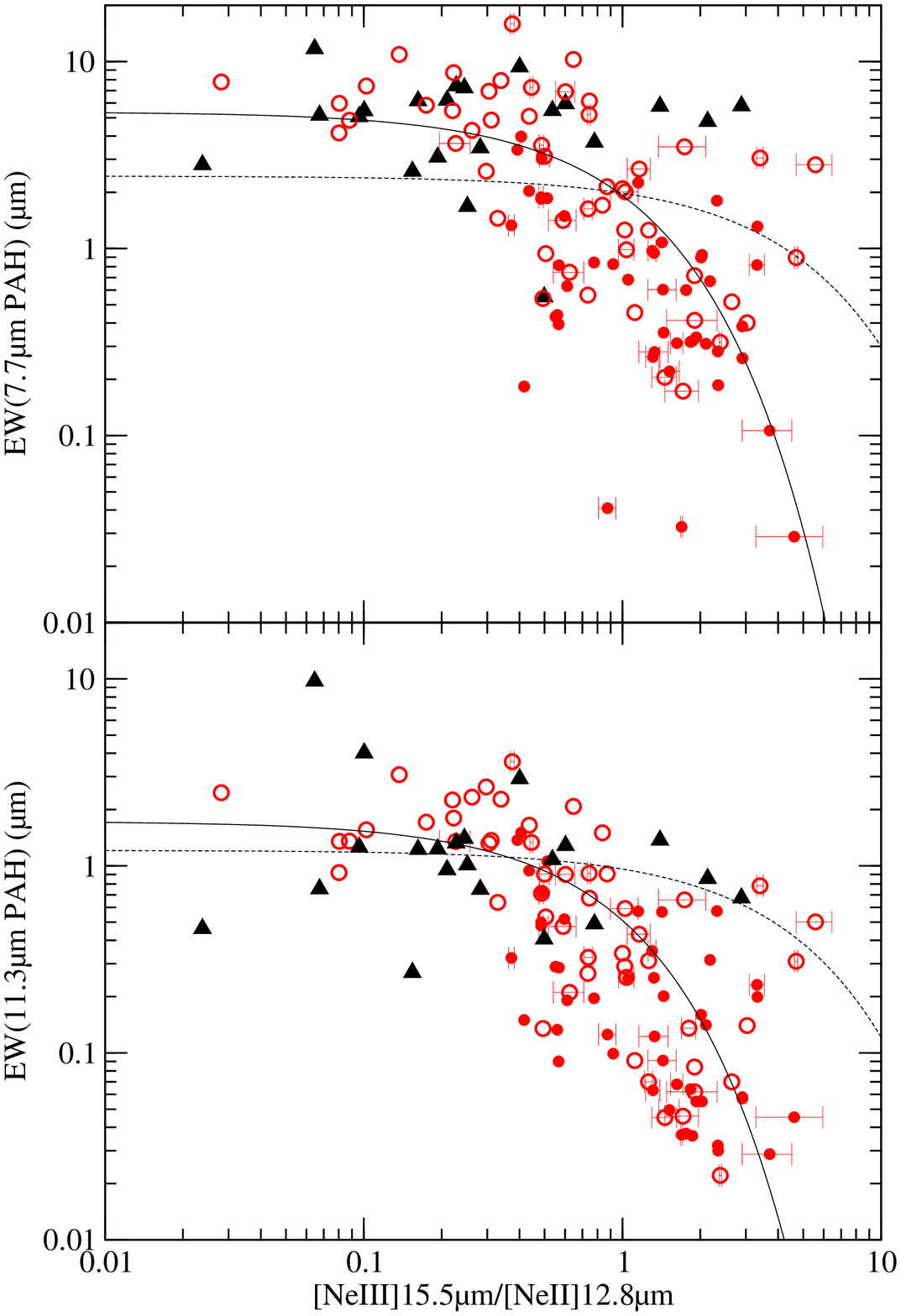}%\tabularnewline
\caption{Diagnostic diagram between an indicator of the hardness of the radiation field, [Ne{\sc\,iii]\,}15.5/[Ne{\sc\,ii]\,}12.8$\mu$m versus EW 7.7$\mu$m (top) and EW 11.3$\mu$m (bottom) PAH emission features. Full triangles are SB taken from \citet{brandl06}, full and empty circles are Sy 1 and Sy 2 of our and \citet{gallimore10} sample. Error bars are shown only for the 15 new AGNs, similar error were found for the remaing objects. The lines represent an exponential regression fitted for the total sample (dotted line) and  for AGNs  sub-sample (solid line).}
\label{ewpahs}
\end{figure*}

In order to test the dependence of the PAH bands strength with the radiation field,  we  compare in  Fig. \ref{ewpahs} the EW of the 7.7$\mu$m and 11.3$\mu$m PAH bands with [Ne{\sc\,iii]\,}/[Ne{\sc\,ii]\,} hardness indicator. It can be seen that  EW of both PAH bands are constant with the radiation field indicator for [Ne{\sc\,iii]\,}/[Ne{\sc\,ii]\,} $\leq$ 0.8, and   fall rapidly with the increasing radiation field. Galaxies, powered by star formation, have a nearly constant EW across the full range of [Ne{\sc\,iii]\,}/[Ne{\sc\,ii]\,}. While for  Sy~1 the EWs of the PAH bands 
decrease rapidly, for the  Sy~2 the EW values are distributed either in the SB or Sy~1 region. Their location in the diagram may reflect the stellar or AGNs dominant component contribution to the spectra (i.e. in some case the Sy~2 may be dominated by a non-thermal component, and in other by circumnuclear massive star formation).  In order to have a more quantitative analysis of 
 Fig. \ref{ewpahs}, we have computed an exponential regression taking into account all objects (AGNs+SBs, dotted line) and another for a  sub-sample formed only by AGNs (solid line). 
 For the latter we have found a correlation coefficient of $\sim$ -0.7 while for the former we have not found any correlation ($\sim$ -0.2).

We conclude this section by arguing that  both  PAH molecules, either ionized (7.7$\mu$m) or neutral (11.3$\mu$m), may be destroyed with increasing hardness of the radiation field.  We emphasize that similar results were found by \citet{genzel98}, which compares the PAH strength with the high to low excitation line ratios for a sample of ULIRS (ultra luminous infrared galaxies) and AGNs. In addition, \citet{baum} has  shown  that the diagram EW PAH 6.2$\mu$m $\times$  [Ne {\sc v}] 14.3$\mu$m/[Ne {\sc ii}] 12.8$\mu$m separate AGNs from SB galaxies.

\section{Summary and Conclusions}\label{conclusions}

We have analyzed the emission line properties of a sample of 186 Spitzer
galaxy spectra.  171 objects were taken from the literature and we have added 15 unpublished AGNs Spitzer
spectra, taken from the public archive. The continuum properties of the 15 new spectra were also studied. Our
main results can be summarized as follows.

\begin{itemize}

\item The normalized Mid-IR continuum spectra (at $\lambda$\,=\,23$\mu$m), of
the 15 new AGNs, steeply rise for redder wavelengths and can be fitted with a warm blackbody
distribution of T $\sim$ 150\,-\,300\,K.

\item  The brightest PAH spectral bands (6.2, 7.7, 8.6, 11.3, and 12.7\,$\mu$m)
and the forbidden emission lines of [Si{\sc\,ii]} 34.8\,$\mu$m, [Ar{\sc\,ii]} 6.9\,$\mu$m, 
[S{\sc\,iii]} 18.7 and 33.4\,$\mu$m were detected in all the SB and in $\sim$80\% of the Sy~2.
 Taking under consideration only the PAH bands at 7.7$\mu$m, 11.3$\mu$m, and 12.7$\mu$m we find they are present in $\sim$80\% of the Seyfert~1, 
while only half of this type of activity show the 6.2$\mu$m  and 8.6 $\mu$m PAH bands.

\item Comparison between observed line intensity ratios of
neutral and ionized PAHs (6.2$\mu$m/7.7$\mu$m $\times$
11.3$\mu$m/7.7$\mu$m) with theoretical models suggests that these ratios, in Sy 1, Sy2 and
SB galaxies, are produced by an appropriate mixture of PAH molecules, with
different sizes and adequate neutral to ionized fraction.

\item  The PAH emission lines observed in Sy 1 and Sy 2 may be due to PAH
molecules with $>$ 180 carbon atoms, while for most of the
low ionization objects (SB and LINERs) it is biased towards molecules
with $<$ 180 carbon atoms. In addition, Seyfert galaxies
appear to be located near to region populated by ionized PAH
molecules, while the majority of SB, H{\sc ii}, and LINER populate the
region of neutral PAHs. These results are consistent with a picture
where Sy 1 and Sy 2 have higher ionization fraction and larger PAH molecules than SB galaxies.

\item We have investigated the PAH line intensity ratio diagrams
6.2$\mu$m/11.3$\mu$m $\times$ 7.7$\mu$m/ 11.3$\mu$m and
6.2$\mu$m/8.6$\mu$m $\times$ 7.7$\mu$m/8.6$\mu$m. 
The separation in activity types (SB, Sy 1 and Sy2), along the vertical
axis in both diagrams, is interpreted as due to the fact that in average the emitting
PAH molecules in Seyferts are larger than those of SB, with ionization fraction and emitting
molecules sizes increasing from SB to AGNs. No separation between Seyfert type is observed.

\item Diagnostic diagram [Ne{\sc\,v]}14.3$\mu$m/[Ne{\sc\,ii]}12.8$\mu$m $\times$
[Ne{\sc\,iii]}15.5$\mu$m/[Ne{\sc\,ii]}12.8$\mu$m
shows that AGNs are correlated, while SB galaxies show a large
scattering. Moreover, for most of the  Sy~1
[Ne{\sc\,iii]\,}/[Ne{\sc\,ii]\,} $\geq$ 0, while for SB and Sy~2s  [Ne{\sc\,iii]\,}/[Ne{\sc\,ii]\,} $\leq$ 0, indicating that both line ratios
can be used as an indicator of radiation field hardness.

\item The ratio between the 7.7$\mu$m and 11.3$\mu$m PAH bands is
nearly constant with the increase of [Ne{\sc\,iii]}15.5$\mu$m/[Ne{\sc\,ii]} 12.8$\mu$m,
indicating that the fraction of ionized to neutral PAH bands does not depend on the hardness of the radiation field.

\item The equivalent width of both PAH features (7.7$\mu$m and
11.3$\mu$m) show the same dependence with the [Ne{\sc\,iii]}/[Ne{\sc\,ii]} ratio. In the case of Sy~1, they
are nearly constant for  [Ne{\sc\,iii]\,}/[Ne{\sc\,ii]\,} $\leq$ 0.8,
and fall sharply with increasing radiation field. For SB galaxies
the PAH EW is constant with the increase
of [Ne{\sc\,iii]}/[Ne{\sc\,ii]}.  Sy~2s are distributed
either in the SB or Sy~1 regions. These results
suggest that the PAH molecules emitting either ionized (7.7$\mu$m) or
neutral (11.3$\mu$m) bands, may be destroyed with increasing hardness of the radiation field.

\end{itemize}

\acknowledgments

We thank an anonymous referee for useful comments, as well as Charles Bonatto, Daniel Ruschel-Dutra and Rogemar Andr\'e Riffel for useful discussions. This research has made use of the NASA/IPAC Extragalactic Database (NED) which is operated by the Jet Propulsion Laboratory, California Institute of Technology, under contract with the National Aeronautics and Space Administration. This work is based on observations made with the Spitzer Space Telescope, which is operated by the Jet Propulsion Laboratory, California Institute of Technology under a contract with NASA.

\begin{table}
\begin{scriptsize}
\renewcommand{\tabcolsep}{0.70mm}
\caption{\label{tabgeral}Sample properties}    
\begin{tabular}{lccc}
\hline\hline
\noalign{\smallskip}
\noalign{\smallskip}
Name	  &	 RA\footnotemark[1] &  DEC\footnotemark[1]  & \textit{z}\footnotemark[1] \\
	  &		&      &	\\
\hline
Sey 1 	&	&	& 	\\
\hline
Mrk279\,	&	13h53m03.4s\,     &     +69d18m30s\,  	  &	   0.030451\,    \\
Mrk334\,  &	00h03m09.6s\,     &     +21d57m37s\,  	  &	   0.021945\,    \\
Mrk478\,	&	14h42m07.4s\,     &     +35d26m23s\,      &    0.079055\,    \\
NGC4748\,	&	12h52m12.4s\,     &     -13d24m53s\,       &	0.014630\,   \\
\hline
Sey 2	&	&	& 	\\
\hline
Mrk3\,	&	06h15m36.3s\,     &     +71d02m15s\,  	  &	    0.013509\,  \\
Mrk471\,	&	14h22m55.4s\,     &     +32d51m03s\,  	  &	    0.034234\, \\
Mrk609\,	&	03h25m25.3s\,     &     -06d08m38s\,  	  &	    0.034488\, \\
Mrk622\,	&	08h07m41.0s\,		&     +39d00m15s\,	  &	    0.023229\, \\
Mrk883\,	&	16h29m52.9s\,     &     +24d26m38s\,  	  &	    0.037496\, \\
Mrk1066\,	&	02h59m58.6s\,     &     +36d49m14s\,  	  &	    0.012025\, \\
NGC1275\,	&	03h19m48.1s\,     &     +41d30m42s\,  	  &	    0.017559\, \\
NGC2622\, &	08h38m10.9s\,     &     +24d53m43s\,  	  &	    0.028624\, \\
NGC3786\,	&	11h39m42.5s\,     &     +31d54m33s\,  	  &	    0.008933\, \\
NGC5728\,	&	14h42m23.9s\,     &     -17d15m11s\,  	  &	    0.009353\, \\
NGC7682\,	&	23h29m03.9s\,     &     +03d32m00s\,  	  &	    0.017140\, \\
\noalign{\smallskip}
\hline
\end{tabular}
\\ \footnotemark[1]{Taken from the NASA Extra-galactic Database (NED).}
\end{scriptsize}
\end{table}

\begin{landscape}
\begin{table}
\renewcommand{\tabcolsep}{.65mm}
\begin{tiny}
\caption{Aromatic Emission Line Strengths ($10^{-16}\,W\,m^{-2}$) and Equivalent Widths ($\mu$m).}
\label{fluxpah}
\begin{minipage}[b]{1.0\linewidth}
\begin{tabular}{lcccccccccccccccccccc}
\noalign{\smallskip}
\hline\hline
Name & 6.2$\mu$m & Ew & 6.7$\mu$m & Ew & 7.7$\mu$m & Ew & 8.3$\mu$m & Ew & 8.6$\mu$m & Ew & 10.7$\mu$m & Ew & 11.3$\mu$m & Ew & 12.0$\mu$m & Ew & 12.7$\mu$m & Ew & 17.0$\mu$m & Ew \\
\noalign{\smallskip}
\hline
\noalign{\smallskip}
\multicolumn{21}{c}{Sy~1} \\
\noalign{\smallskip}
\hline
Mrk279   &     2.88$\pm$0.33  &  0.035      &    7.19$\pm$0.81&	 0.095 &     15.6$\pm$   1.82    &	 0.220& 	  -		   &	  -    &	 0.09$\pm$0.26& 	0.001&      0.84$\pm$   0.12&     0.013   &      2.94$\pm$	0.36&	  0.049&	   -		   &	   -&	  0.15$\pm$0.10  &	0.002	&     6.62$\pm$0.89 &    0.154\\
Mrk334   &     28.5$\pm$0.30  &   0.548    &    9.92$\pm$0.67&	 0.183 &     107$\pm$2.88    &	  1.881&       7.98$\pm$0.34&	 0.138& 	   20.2$\pm$0.24&	  0.351&       1.12$\pm$   0.09&	0.019  &       28.3$\pm$    0.35&	  0.479&	8.05$\pm$   0.19&   0.131   &     16.7$\pm$   0.39&	 0.263 &   18.5$\pm$	  0.83 &    0.229\\
\hline
\noalign{\smallskip}
\multicolumn{21}{c}{Sy~2} \\
\noalign{\smallskip}
\hline
Mrk3     &   0.77$\pm$0.31  & 0.006	  &    4.56$\pm$0.63&     0.037 &	 38.2 $\pm$	2.98 &         0.315&	       -		&	-    &  	 1.14$\pm$0.35&        0.009&	     1.22$\pm$     0.17&    0.009 &      3.28$\pm$   0.55&    0.022&	     1.61$\pm$   0.37&    0.009   &      1.45$\pm$ 0.16&      0.008 &	       39.2$\pm$1.57 &    0.143\\
Mrk471   & 2.67$\pm$0.12  &    0.820   &    3.98$\pm$0.32&	     1.030 &	  6.78 $\pm$1.15&  	1.415&  	-		&	-    &  	 2.15$\pm$0.09& 	0.382&      0.05$\pm$ 0.04&    0.007    &      3.30$\pm$0.22   &	0.474&       0.56$\pm$   0.19&     0.078  &      1.67$\pm$0.29& 	0.226 &      5.18$\pm$0.26 &    0.359\\
Mrk609   &    16.9$\pm$0.22  &    1.020   &    8.57 $\pm$0.49&      0.485  &   58.0 $\pm$1.74&  	3.115&       4.62$\pm$  0.20&	 0.245& 	11.7$\pm$0.16&  	0.617&      0.53$\pm$  0.07&     0.027  &      17.7$\pm$   0.32&	0.903&        5.76$\pm$   0.23&   0.282&	10.4$\pm$   0.48  &         0.474  &       16.5$\pm$	0.66 &    0.469\\
Mrk622   &   5.01 $\pm$0.19  &   0.516   &    2.26$\pm$0.42&      0.194 &	 14.6 $\pm$	1.65 &  	0.986&  	-		&	-    &  	 1.71$\pm$0.14&        0.093&	   0.37$\pm$0.07&     0.015    &       6.45$\pm$  0.28  &     0.253&	       2.40$\pm$   0.18&  0.089   &      4.41$\pm$ 0.41&       0.154 &       2.65$\pm$0.49 &   0.044\\
Mrk883   &    4.62$\pm$0.14  &  1.100	  &    4.68$\pm$0.35&     0.920 &	 10.7 $\pm$	1.18 &  	1.630&  	-		 &	-    &  	 1.33$\pm$0.10& 	0.161&    0.01$\pm$0.04&   0.001       &       3.76$\pm$  0.22&      0.324&	     0.83$\pm$   0.18&     0.068&	 1.37$\pm$  0.34  &         0.107  &        2.53$\pm$	0.30 &     0.111\\
Mrk1066  &       47.0$\pm$0.28  &   0.757   &	 15.5$\pm$0.58&	0.263 &     199 $\pm$2.76&  	3.567&       18.6$\pm$0.37&    0.340   &  	 31.9$\pm$0.26& 	0.592&       1.29$\pm$     0.11&	0.020&       52.1$\pm$    0.43&	0.713&        15.3$\pm$   0.23&   0.181   &     30.9$\pm$0.45&        0.328 &        43.0$\pm$	 1.41 &    0.325\\
NGC1275  &    10.8$\pm$0.25  &  0.097	 &	18$\pm$0.48&      0.139 &	     -  	     &  	   - &         -		&      -     &  	   -	      & 	   -&	    10.6$\pm$  0.10&      0.043 &      17.4$\pm$ 0.33&       0.070&	      2.90$\pm$   0.22&     0.011 &         -		&	   - &       53.9$\pm$0.98 &    0.199\\
NGC2622  &    2.75$\pm$0.14  &   0.233   &    4.33$\pm$0.37&      0.339 &		    -	     &       -       &  	-		 &	-   &		    -	      & 	    -&      0.32$\pm$ 0.05&     0.015   &      2.97$\pm$0.21  &     0.135&    0.78$\pm$    0.14    &     0.034     &      2.03$\pm$ 0.13&       0.086  &       2.42$\pm$	 0.46 &   0.0472\\
NGC5728  &   10.6$\pm$0.32  &	0.379	&    4.11$\pm$0.64&      0.167 &	 65.1 $\pm$	3.00 &  	 3.050&      4.88$\pm$0.37&      0.245&		 8.09$\pm$0.31& 	0.424&      0.04$\pm$  0.21&    0.001   &      21.4$\pm$   0.65&	0.783&        3.97$\pm$   0.35&   0.117   &      9.89$\pm$ 0.78&       0.258 &       32.6$\pm$ 1.14 &    0.471\\
NGC7682  &     40.4$\pm$0.16  &    1.840   &	11.5$\pm$0.35&      0.456 &     160 $\pm$1.48&  	5.200&       11.3$\pm$0.16&      0.328&		 29.7$\pm$0.12& 	0.828&       1.07$\pm$   0.05&     0.024  &       40.9$\pm$  0.19  &	0.913&        10.1$\pm$   0.12&   0.221   &     24.0$\pm$0.26&         0.522 &       39.8$\pm$0.55		&    0.641\\
\hline
\end{tabular}
\end{minipage}
\caption{Atomic Emission Line Strengths ($10^{-16}\,W\,m^{-2}$) and Equivalent Widths ($\mu$m).}
\label{fluxionic}
\begin{minipage}[b]{1.0\linewidth}
\begin{tabular}{lcccccccccccccccccccc}
\noalign{\smallskip}
\hline\hline
Name & [Ar{\sc\,ii]\,}6.9$\mu$m & Ew & [Ar{\sc\,iii]\,}8.9$\mu$m & Ew &[S{\sc\,iv]\,}10.5$\mu$m & Ew & [Ne{\sc\,ii]\,}12.8$\mu$m & Ew &
[Ne{\sc\,iii]\,}15.5$\mu$m & Ew & [S{\sc\,iii]\,}18.7$\mu$m & Ew & [O{\sc\,iv]\,}25.9$\mu$m & Ew & [Fe{\sc\,ii]\,}25.9$\mu$m & Ew & [S{\sc\,iii]\,}33.4$\mu$m & Ew &
[Si{\sc\,ii]\,}34.8$\mu$m & Ew \\
\hline
\noalign{\smallskip}
\multicolumn{21}{c}{Sy~1} \\
\noalign{\smallskip}
\hline
Mrk279 &     0.70$\pm$0.16&    0.014&	  0.05$\pm$ 0.10&    0.001&	  0.60$\pm$0.05&     0.014 &	  0.82$\pm$	0.05&    0.022 &	 1.26 $\pm$	0.07&     0.039&	0.64 $\pm$0.10&	  0.024&	 1.10$\pm$0.05&	 0.067&		  -		 &	      -&	 -		&	     -&      0.87$\pm$0.13&     0.088\\
Mrk334 &      1.36$\pm$0.09&    0.036&	  0.06$\pm$0.05&     0.001&	  0.63$\pm$0.03&     0.016 &	   5.32$\pm$	0.03&     0.122 &	 2.58$\pm$0.04&     0.050&		 2.91$\pm$0.06&	    0.050&	   -		&	      -&      0.49$\pm$0.06 &	0.008&      3.82$\pm$0.14&	0.068&       7.07$\pm$0.14&      0.129\\
Mrk478	&  0.47 $\pm$0.03   &   0.012   &   0.06  $\pm$	0.06   &     0.002  &	0.38$\pm$0.04  &   0.016 &  0.59      $\pm$	0.04   &    0.032  &   0.52 $\pm$0.01   &	   0.032  &   0.40$\pm$0.01     &   0.027   &   0.75 $\pm$0.02    &    0.066  &   -				    &	-    &    -	$\pm$-	&    -      &	0.01 $\pm$0.06    &    0.001\\
NGC4748	&  0.45 $\pm$0.21   &   0.030   &   0.28  $\pm$	0.12   &     0.018  &	2.34$\pm$0.09  &   0.144 &  1.14      $\pm$	0.06   &    0.065  &   3.78 $\pm$0.12   &	   0.123  &   1.83$\pm$0.15     &   0.058   &   6.81 $\pm$0.13    &    0.269  &   -				    &	-    &    2.69  $\pm$	0.24	&   0.131   &	3.42 $\pm$0.29    &    0.167\\
\hline
\noalign{\smallskip}
\multicolumn{21}{c}{Sy~2} \\
\noalign{\smallskip}
\hline
Mrk3 &      3.31$\pm$0.19&    0.040&	    2.63$\pm$ 0.12&      0.034&	   6.26$\pm$0.08&     0.072 &	   8.75$\pm$	0.08&    0.073 &	 20.9 $\pm$	0.12&      0.111&	  6.70 $\pm$	0.11&	  0.037&	18.2$\pm$0.10&	  0.159&	  -		 &	      -&      4.25$\pm$0.20&	0.054&       11.7$\pm$0.20&      0.156\\
Mrk471 &      0.43$\pm$0.06& 0.153	&	-		&	   -&	  0.12$\pm$0.01&     0.028 &	  0.32$\pm$	0.02&    0.064 &	0.19$\pm$0.01&       0.020&	0.13$\pm$0.02&	  0.013&      0.36$\pm$0.02&	 0.049&	  -			 &	      -&      1.74$\pm$0.03&	 0.285&       2.03$\pm$0.08&      0.331\\
Mrk609 &      1.02$\pm$0.08& 0.083	&		-	&	  -&	  0.29$\pm$0.02&     0.022 &	   2.22$\pm$	0.03&     0.144 &	 1.11$\pm$0.04&     0.047&		 1.19$\pm$0.05&	  0.048&      0.82$\pm$0.03&	 0.034&		  -		&	      -&      1.59$\pm$0.08&	0.070&       3.63$\pm$0.13&      0.159\\
Mrk622 &     0.40$\pm$0.08&    0.046&	      -			&	   -&	  0.34$\pm$0.02&     0.021 &	  0.70$\pm$	0.03&    0.036 &	0.73$\pm$0.03&      0.019&		0.39$\pm$0.07&	 0.009&      0.73$\pm$0.04&		 0.020&	  -			 &	      -&     0.80$\pm$0.09&	 0.028&       3.51$\pm$0.14&      0.128\\
Mrk883 &     0.55$\pm$0.06& 0.147	&0.11$\pm$0.02&     0.019&	  0.26$\pm$0.02&     0.036 &	   1.31$\pm$0.02&      0.150 &	0.96$\pm$0.03&     0.069&		 0.71$\pm$0.03&	  0.043&      0.91$\pm$0.02&	 0.058&		  -		 &	      -&     0.82$\pm$0.05&	0.052&       2.58$\pm$0.09&      0.161\\
Mrk1066 &      3.75$\pm$0.12&    0.096&0.83$\pm$0.07&     0.023&	   1.61$\pm$0.04&     0.038 &	    9.60$\pm$	0.03&     0.146 &	 4.67 $\pm$	0.09&     0.055&	  4.80$\pm$0.12&	  0.050&	3.51$\pm$0.22&	 0.031&      0.25$\pm$ 0.29 &	0.002  &      3.72$\pm$0.22&	0.035&       11.2$\pm$0.23&      0.108\\
NGC1275 &      1.89$\pm$0.11&    0.019&	       -		&	   -&	   2.78$\pm$0.05&      0.017 &	   3.37$\pm$	0.05&    0.019 &	 4.25 $\pm$	0.14&     0.023&	 1.77 $\pm$	0.11&	 0.009&		   -		&	      -&	  -		 &	      -&    0.07$\pm$0.18&	0.001&       10.2$\pm$0.23&     0.096\\
NGC3786 &      0.56$\pm$0.06& 0.031	&	-	&	  -&	  0.51$\pm$0.02&     0.031 &	    1.49$\pm$	0.02&     0.087 &   1.10$\pm$0.02&	  0.042& 0.06$\pm$0.03&	 0.024&	  1.31$\pm$0.19 &	0.066&      0.04 $\pm$0.01& 0.020&    1.33$\pm$0.06&    0.076 & 1.70$\pm$0.05 & 0.096\\
NGC2622 &     0.62$\pm$0.05&    0.069&	   0.11$\pm$0.04&    0.009&	  0.44$\pm$0.02&     0.031 &	  0.50$\pm$	0.02&    0.031 &	0.91$\pm$0.03&     0.025&		0.40$\pm$0.04&	  0.012&      0.86$\pm$0.06&	  0.036&      0.23$\pm$0.05 &	0.009  &     0.59$\pm$0.05&	 0.036&       2.56$\pm$0.08&      0.165\\
NGC5728 &     0.95$\pm$0.10&    0.061&	   0.42$\pm$0.09&     0.035&	   2.84$\pm$0.10&      0.197 &	   2.26$\pm$	0.06&    0.084 &	 7.69$\pm$0.06&      0.162&	 3.85$\pm$0.08&	  0.081&	13.7$\pm$0.09&	  0.244&	  -		 &	      -&      6.29$\pm$0.12&	0.094&       9.87$\pm$0.17&      0.145\\
NGC7682 &      1.34$\pm$0.05& 0.073	&	-		&	  -&	  0.86$\pm$0.02&     0.029 &	    5.10$\pm$	0.02&     0.164 &	 3.79$\pm$0.04&      0.109&	  4.60$\pm$0.05&	  0.093&	4.42$\pm$0.04&	 0.077&	  -			 &	      -&      6.03$\pm$0.09&	 0.102&       12.2$\pm$0.11&      0.203\\
\hline
\end{tabular}
\end{minipage}
\end{tiny}
\end{table}
\end{landscape}

\end{document}